# Echo state graph neural networks with analogue random resistor arrays

Shaocong Wang[†,1,2,3], Yi Li[†,2,4], Dingchen Wang[1,3], Woyu Zhang[2,4], Xi Chen[1,3], Danian Dong[2,4], Songqi Wang[2], Xumeng Zhang[5], Peng Lin[6], Claudio Gallicchio[7], Xiaoxin Xu[2,4], Qi Liu[2,5], Kwang-Ting Cheng[3,8], Zhongrui Wang[*,1,3], Dashan Shang[*,2,4], Ming Liu[2,5]

[1]Department of Electrical and Electronic Engineering, the University of Hong Kong, Hong Kong, China

[2]Key Laboratory of Microelectronic Devices & Integrated Technology, Institute of Microelectronics, Chinese Academy of Sciences, Beijing 100029, China

[3]ACCESS – AI Chip Center for Emerging Smart Systems, InnoHK Centers, Hong Kong Science Park, Hong Kong, China

[4]University of Chinese Academy of Sciences, Beijing 100049, China

[5]Frontier Institute of Chip and System, Fudan University, Shanghai 200433, China

[6]College of Computer Science and Technology, Zhejiang University, Zhejiang, 310027, China

[7]Department of Computer Science, University of Pisa, Largo B. Pontecorvo, 3. 56127 Pisa, Italy

[8]Department of Electronic and Computer Engineering, the Hong Kong University of Science and Technology, Hong Kong, China

[†] These authors contributed equally.

[*] e-mails: zrwang@eee.hku.hk; shangdashan@ime.ac.cn

## Abstract

Recent years have witnessed an unprecedented surge of interest, from social networks to drug discovery, in learning representations of graph-structured data. However, graph neural networks, the machine learning models for handling graph-structured data, face significant challenges when running on conventional digital hardware, including von Neumann bottleneck incurred by physically separated memory and processing units, slowdown of Moore's law due to transistor scaling limit, and expensive training cost. Here we present a novel hardware-software co-design, the random resistor array-based echo state graph neural network, which addresses these challenges. The random resistor arrays not only harness low-cost, nanoscale and stackable resistors for highly efficient in-memory computing using simple physical laws, but also leverage the intrinsic stochasticity of dielectric breakdown to implement random projections in hardware for an echo state network that effectively minimizes the training cost thanks to its fixed and random weights. The system demonstrates state-of-the-art performance on both graph classification using the MUTAG and COLLAB datasets and node classification using the CORA dataset, achieving 34.2×, 93.2×, and 570.4× improvement of energy efficiency and 98.27%, 99.46%, and 95.12% reduction of training cost compared to conventional graph learning on digital hardware, respectively, which may pave the way for the next generation AI system for graph learning.

**Introduction**

The great success of graph neural networks[1,2], graph convolutional networks[3], and graph attention networks[4] well illustrates the power of machine learning in handling graph-structured data which simultaneously characterize both objects and their relationships. As a result, graph learning[5] is quickly standing out in many real-world applications such as prediction of chemical properties of molecules for drug discovery[6], recommender systems of social networks[7], and combinatorial optimization for design automation[8]. In the era of big data and Internet of Things (IoT), the size and scale of graph-structured data are exploding. For example, the social network of Facebook comprises more than two billion users and one trillion edges representing social connections.[9] This imposes a critical challenge to the current graph learning paradigm which implements graph neural networks on conventional complementary metal-oxide-semiconductor (CMOS) digital circuits. Such digital hardware suffers from frequent and massive data shuttling between off-chip memory and processing units during graph learning, the so called von Neumann bottleneck.[10-19] Furthermore, the technology node of transistors has reached 3 nm, just a few unit cells of silicon. This leads to inevitable slowdown of Moore's Law which has fuelled the past development of CMOS chips in the last few decades. As a result, further scaling of transistor is becoming less cost-effective. Last but not least, the training of graph neural networks is expensive, due to tedious error backpropagation for node and graph embedding. For example, the training of PinSage took 78 hours on 32 central processing unit (CPU) cores and 16 Tesla K80 graphic processing units (GPUs)[20]. The growing challenges in both hardware, *i.e.* von Neumann bottleneck and transistor scaling, as well as software *i.e.* tedious training, calls for a brand-new paradigm of graph learning.

Resistive memory may provide a hardware solution to these issues[21-45]. When these resistive elements are grouped into a crossbar array, they can naturally perform vector-matrix multiplication, one of the most expensive and frequent operations in graph learning[46,47]. The matrix is stored as the conductance of the resistive memory array, where Ohm's law and Kirchhoff's current law physically govern the multiplication and summation, respectively. As a result, the data is both stored and processed in the same location. This in-memory computing concept can largely obviate the energy and time overheads incurred by expensive off-chip memory access in graph learning on conventional digital hardware. In addition, the resistive memory cells are of simple capacitor-like structures, equipping them with excellent scalability and 3D stack-ability. However, resistive memory suffers from a series of issues when changing their resistance, that can defeat the efficiency advantage of in-memory graph learning that demands frequent updates of resistive weights. This is because resistive memory relies on electrochemical reactions or phase changes to adjust their conductance[48-53]. This mechanism results in switching energy and duration that are orders of magnitude higher than those of transistors. In addition, the inevitable stochasticity associated with ionic or atomic motions makes precise resistance changes difficult. As a result, graph learning has not yet experimentally leveraged the advantage of resistive in-memory computing.

Here, we propose a novel hardware-software co-design, the random resistor array-based echo state graph neural networks (ESGNN)[54,55]. The marriage of random resistor arrays and ESGNN not only retains the boost of energy-area efficiency thanks to in-memory computing on random resistors, but also makes use of intrinsic stochasticity of dielectric breakdown to provide low-cost and nanoscale hardware embodiment of ESGNN. Moreover, the echo state network employs iterative random projections for node and graph embedding, which gets rid of the tedious training of conventional graph neural networks, enabling efficient and affordable real-time graph learning.

In this article, we showcase such a co-designed ESGNN physically implemented on a 40 nm computing-in-memory macro to accelerate graph learning. We demonstrate state-of-the-art graph classification results on the MUTAG and COLLAB datasets, as well as node classification on the CORA dataset. We observe a 34.2×, 93.2×, and 570.4× improvement in energy efficiency using the random resistor matrices compared to that of digital alternatives for classifying the MUTAG, COLLAB, and CORA dataset. Furthermore, the training cost is reduced by 98.27%, 99.46%, and 95.12% respectively, thanks to the representation extraction using random projections in ESGNN. Our system paves the way for future efficient and fast graph learning.

# Hardware-software co-design for ESGNN using random resistor arrays

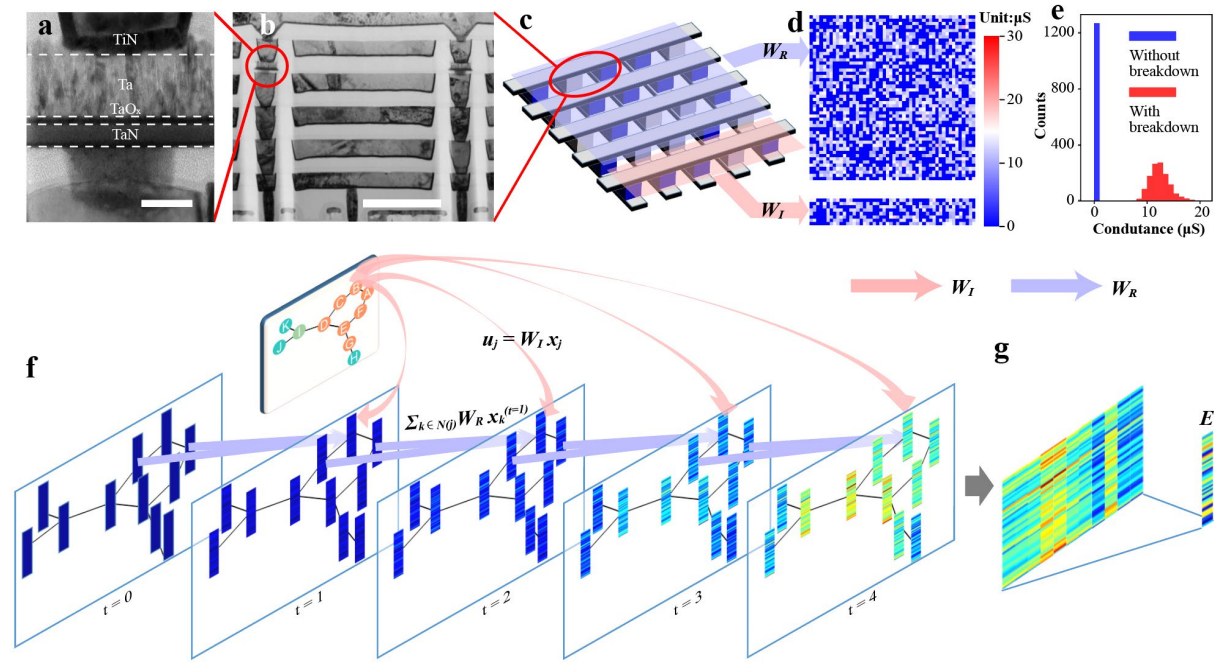

**Figure 1 | Hardware-software co-design of random resistor array-based ESGNN for graph learning.**
**a**, Cross-sectional transmission electron micrograph (TEM) of a single resistive memory cell which works as a random resistor after a dielectric breakdown. Scalebar 20 nm. **b**, Cross-sectional TEM of the resistive memory crossbar array fabricated using the backend-of-line process on a 40 nm technology node tape-out. Scalebar 500 nm. **c**, Schematic illustration of the partition of the random resistor crossbar array, where cells shadowed in blue are the weights of the recursive matrix (passing messages along edges) while those in red are weights of the input matrix (transforming node input features). **d**, Corresponding conductance map of the two random resistor arrays in **c**. **e**, Conductance distribution of the random resistor arrays. **f**, Node embedding procedure of the proposed ESGNN. The internal state of each node at the next time step is co-determined by the sum of neighbouring contributions (blue arrows indicate multiplications between node internal state vectors and the recursive matrix in **d**), the input feature of the node after a random projection (red arrows indicate multiplications between input node feature vectors with the input matrix in **d**), as well as the node internal state in the previous time step. **g**, Graph embedding based on node embeddings. The graph embedding vector is the sum pooling of all node internal state vectors in the last time step.

**Figure 1** illustrates the hardware-software co-design scheme, where random resistor arrays are used to physically implement the ESGNN. Hardware-wise, the random distribution of dielectric breakdown voltages, due to inevitable process variation and motion of ions, provides a natural source of the randomness entropy to produce large-scale random resistor arrays that have been validated for implementations of true random number generators[56] and physically unclonable functions[57]. Here, we fabricated CMOS-compatible nanoscale $TaN/TaO_x/Ta/TiN$ resistive memory cells (**Figure 1a** and **Figure 1b**) in crossbar arrays (**Figure 1c**) using the backend-of-line process on a 40 nm technology node tape-out (see Methods). The chip is integrated on a printed circuit board with analogue-digital conversion circuitry and a Xilinx ZYNQ system-on-chip (SoC), constituting a hybrid analogue-digital computing platform (see Method and Supplementary Figure 1 for the system design and photo). As shown in **Figure 1c**, the crossbar array is then partitioned into two sub-arrays to represent two weight matrices, $\boldsymbol{W_I} \in \mathbb{R}^{h \times (u+1)}$ the input matrix and $\boldsymbol{W_R} \in \mathbb{R}^{h \times h}$ the recursive matrix, of the ESGNN, where $u$ and $h$ represent the input dimension and the hidden dimension of each node, respectively (see Methods on mapping resistor conductance to weights). Biasing all cells of as-deposited resistive memory to the median of their breakdown voltages, some cells will experience dielectric breakdown if their breakdown voltages are lower than the applied voltage, forming random resistor arrays as illustrated by the conductance maps of both $\boldsymbol{W_I}$ and $\boldsymbol{W_R}$ in **Figure 1d** (see Supplementary Figure 2 for the stochasticity of dielectric breakdown voltages). Compared to pseudo random number generation using digital systems, the source of randomness here is the stochastic redox reactions and ion migrations arising from compositional inhomogeneity of resistive memory cells, offering low-cost and highly scalable random resistor arrays for in-memory computing. The corresponding histogram in **Figure 1e** shows a conductance gap between breakdown cells and those remain insulating. The conductance distribution of the former follows a stable quasi-normal distribution, which can be tailored by both fabrication conditions and electrical operation parameters, offering tuneable hardware implementations of ESGNN (see Supplementary Figure 3 for the stability of the random resistor arrays).

Software-wise, the echo state network is a type of reservoir computer[26,31,43,58] comprising a large number of neurons with random and recurrent interconnections, where the states of all neurons are accessible by a simple software readout layer[55,59]. The consecutive nonlinear random projections in the high-dimensional state space produce trajectories at the edge of chaos, benefiting graph embedding extraction[55]. Here the network parameters of ESGNN are physically embodied by the two random resistor arrays, where the input matrix $W_I$ modulates the influence of a node input feature vector on the node internal state, and the recursive matrix $W_R$ determines the influence of neighbouring nodes on the same node internal state (see Methods for the choice of resistance scaling factors to ensure the echo state property). This graph embedding process is schematically illustrated in **Figure 1f**. For a given graph, the initial embedding of node $j$ is a zero vector $s_j^{(0)} \in \mathbb{R}^h$. The input feature of the same node $x_i \in \mathbb{R}^{u+1}$ will first undergo a random projection using the input matrix to produce its input projection $u_j = W_I x_j \in \mathbb{R}^h$. In each subsequent time step, a new state of the node is computed by aggregating its current state $s_j^{(t)}$, input projection $u_j$, and the states of all its neighbors after random projections by the recursive matrix $\sum_{k \in N(j)} W_R s_k^{(t)}$ (here $N(j)$ denotes the set of neighbouring nodes of node $j$). These consecutive random projections to high-dimensional space paired with nonlinear activations endow each node with a unique and discriminative representation. The final internal states of nodes, or node embeddings, will be used to create a graph embedding $g \in \mathbb{R}^h$ by sum pooling for graph classification problems, as illustrated in **Figure 1g** (see Methods for the details of node and graph embedding).

**Graph classification with ESGNN**

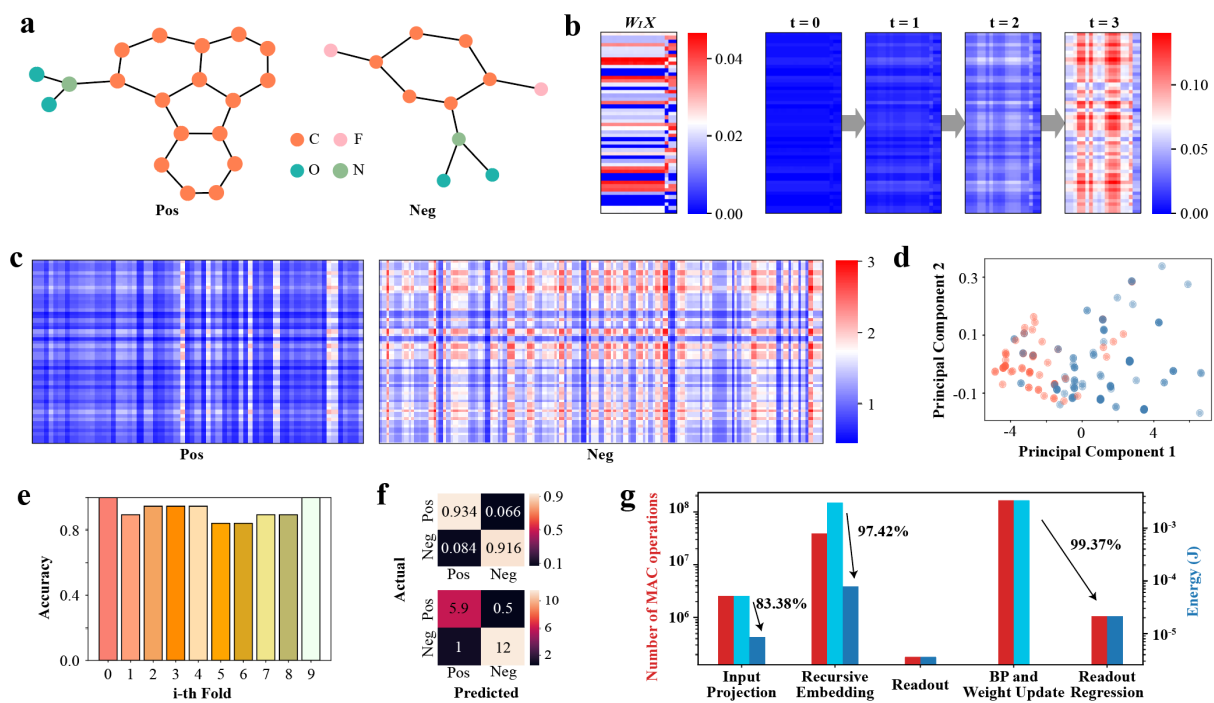

**Figure 2 | Classification of molecular graphs. a**, Illustration of samples from MUTAG molecular dataset, where nodes of different colours represent different atoms while edges are chemical bonds. Depending on the mutagenicity, these molecules are categorized into positive and negative classes. **b**, An example of MUTAG node embedding process. The input features of nodes are first projected onto the state space using the input matrix $W_I$, the hidden state of each node is updated according to the protocol shown in **Figure 1f** and Method, which leads to node embeddings that encapsulate graph information. **c**, Graph embedding vectors of the two categories of MUTAG dataset. Each column vector is a graph embedding. The embeddings of the left (right) colormap are from the positive (negative) class. **d**, The graph embeddings are mapped to a 2D space using principal component analysis (PCA). Dots of yellow (blue) colour represent molecules of the positive (negative) mutagenicity, which can be linearly separated. **e**, Accuracy of each fold in a ten-fold cross-validation. The average accuracy is 92.11%, comparable to state-of-the-art algorithms. **f**, Confusion matrices of the experimental classification results. The upper matrix is a ten-fold averaged confusion matrix, which is then horizontally normalized to produce the lower matrix. **g**, Breakdown of estimated multiply and accumulation (MAC) operations (red bars) and associated energy (light blue bars for the conventional digital system and dark blue bars for our system). In forward (backward) propagation, the conventional digital hardware and our random resistor array-based system consumes ~3.08 mJ (and ~2.40mJ) and ~0.09 mJ (~21.17uJ), respectively, revealing >34.2× improvement of the inference energy efficiency (~98.27% reduction of the total training energy cost). BP, backpropagation.

We first solve a representative graph classification task using the MUTAG molecular dataset[60] with a random resistor array-based ESGNN. MUTAG is a widely used molecular dataset that comprises 188 nitroaromatic compounds (see examples in **Figure 2a**). These molecular compounds are essentially graphs. Their nodes stand for atoms while edges denote chemical bonds. The molecular graphs can be divided into two categories according to their mutagenicity, the ability to mutate genes of certain bacteria. **Figure 2b** shows the experimental input projection and evolution of node internal states of a single molecular graph according to the process shown in **Figure 1f**. Internal states of nodes (columns) gradually differ from projected input features by accumulating messages from neighbouring nodes, and thus encode structural information pertaining to the topology of the graph. Here the embedding process is iterated four times, a balance between capturing more topological information and over-smoothing[61]. The final graph embeddings of the entire dataset are shown in **Figure 2c**, in which embeddings of the same class are similar while those from different classes have large contrast. **Figure 2d** visualizes distribution of graph embeddings by mapping them to a 2D space via principal component analysis (PCA), where orange and blue dots are graphs from positive and negative classes, respectively (see Supplementary Figure 4 for visualizing graph embedding distributions in 3D space). Although a few samples are mixed on the classification boundary, the majority of samples can be well classified by a simple linear classifier thanks to the dynamics of the echo state network. These embedding vectors will be classified by a simple software readout layer (with 102 floating point weights) optimized by linear regression at low hardware and energy cost (see Methods for the implementation and training of the readout layer). To evaluate the classification performance of the MUTAG dataset, we use ten-fold cross-validation, where the accuracy of each fold is shown in **Figure 2e**. The overall performance of our implementation is ~92.11%, which is comparable to state-of-the-art algorithms such as DDGK (91.58%)[62] and Patchy-SAN (92.60%)[63] running on digital computers. Confusion matrices in **Figure 2f** show, on average, 5.9 out of 6.4 (12 out of 13) molecule compounds of positive (negative) mutagenicity are correctly classified, which translates to a class-wise recognition rate of 91.64% (93.35%) (see Supplementary Figure 5 for confusion matrices of all folds and Supplementary Figure 6 for the impact of hyperparameters on accuracy). To verify the boost of energy efficiency, we conducted a preliminary comparison of energy consumption between the conventional digital hardware and our random resistor array-based system. As shown in **Figure 2g**, the red bars are the breakdown of number of multiply and accumulation (MAC) operations in the ESGNN, while the light blue bars and dark blue bars are the associated estimation of energy consumption of conventional digital hardware and our random resistor array-based system, respectively. Since multiplications with the input matrix $W_I$ and recursive matrix $W_R$ account for the majority of MAC operations and thus power consumption in the forward propagation of the ESGNN, the overall energy of the random resistor array-based system is ~0.09 mJ per forward propagation of the entire dataset compared to ~3.08 mJ of the conventional digital system, resulting in ~34.2× improvement in inference energy efficiency (see Methods and Supplementary Figure 7 for the impact of hyperparameters on system energy efficiency). In addition, as pseudo inverse is used to train the simple readout layer of the ESGNN, the total number of MAC operations (energy) is ~1.0 MOPs (~21.17 µJ), in contrast to ~165.1 MOPs (~3.34 mJ) per epoch of stochastic gradient descent (SGD) for a graph neural network with the same number of weights, leading to a ~98.27% reduction of the total training cost (see Supplementary Note 1).

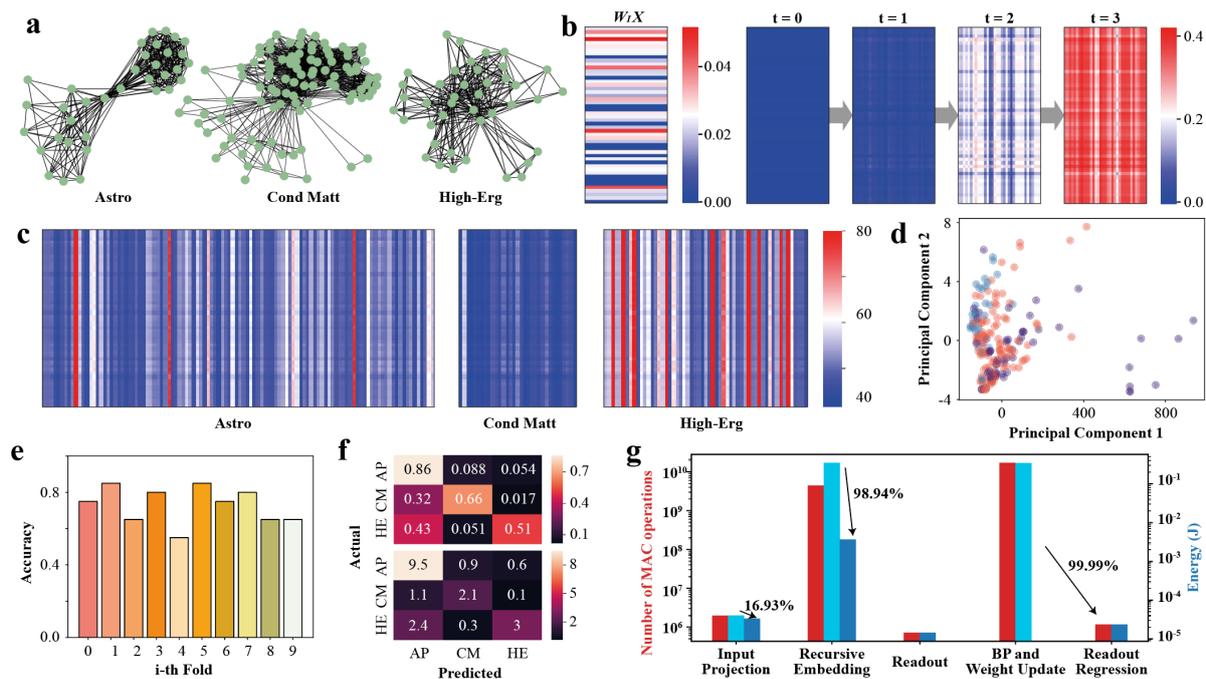

**Figure 3 | Classification of collaboration networks. a**, Example collaboration network graphs from COLLAB dataset that correspond to different branches of physics (e. g. astrophysics or AP, high-energy physics or HE, condense matter physics or CM). Each node denotes a researcher while an edge represents a collaboration relation. **b**, An example of COLLAB node embedding process according to the protocol shown in **Figure 1f** and Methods, which leads to node embeddings that encapsulate more graph information. **c**, Graph embedding vectors of the three categories of COLLAB dataset. Each column is a graph embedding. **d**, Graph embedding mapped to a 2D space using PCA. Orange, blue, and purple dots denote collaboration networks from the AP, CM and HE communities, respectively, which show a clear boundary between AP and CM. **e**, Accuracy of each fold in a ten-fold cross-validation. The average accuracy is 73.00%, comparable to state-of-the-art algorithms. **f**, Confusion matrices of the experimental classification results. The upper matrix is a ten-fold averaged confusion matrix, which is then horizontally normalized to produce the lower matrix. **g**, Breakdown of estimated MAC operations (red bars) and associated energy (light blue bars for the conventional digital system and dark blue bars for our system). In forward (backward) propagation, the conventional digital hardware and our random resistor array-based system consumes ~346.20 mJ (0.43 J) and ~3.72 mJ (23.49 μJ), respectively, revealing >93.2× improvement of the inference energy efficiency (~99.46% reduction of the total training energy cost).

In addition to modelling molecules, the random resistor array-based ESGNN has also been used to solve a representative social network classification problem using the COLLAB dataset[64]. As shown in **Figure 3a**, each graph of the COLLAB dataset depicts a research collaboration network from one of the three branches of physics, the astrophysics, the high energy physics, and the condensed matter physics. Here nodes are researchers while edges denote collaboration relations. We randomly pick 200 graphs from the COLLAB dataset for learning. As nodes, or researchers, in the COLLAB dataset share a unity input feature, rendering input projections of different nodes identical in **Figure 3b**. However, thanks to iterative message passing in ESGNN, node internal states have gradually integrated graph information such as topology along iterations, yielding unique node embeddings shown in the last time step of **Figure 3b**. The final graph embeddings, grouped by classes, are shown in **Figure 3c**, where graphs from condensed matter community and high energy community are well separated from each other due to clear differences in topologies. This is also corroborated by the distribution of graph embedding vectors by mapping them to a 2D space using PCA, as shown in **Figure 3d**, where blue (condensed matter physics) and purple (high energy physics) dots are linearly separable (see Supplementary Figure 4 for visualizing graph embedding distributions in 3D space). On the other hand, graphs from the astrophysics community tend to share similar topologies with the other two, which is also revealed by the fact that pink dots (astrophysics) partially overlap with blue and purple dots. **Figure 3e** shows the classification performance of a ten-fold cross-validation (see Supplementary Figure 5 for confusion matrices of all folds). The random resistor array-based ESGNN is able to achieve state-of-the-art accuracy of 73.00%, compared to GraphSage (73.90%)[65] and DGCNN (73.76%)[66]. **Figure 3f** shows the experimentally acquired confusion matrix of the ten-fold cross-validation (see Supplementary Figure 5 for confusion matrices of all folds and Supplementary Figure 6 for the impact of hyperparameters on accuracy). The accuracy of correctly classifying the astrophysics reaches 85.82%, but 31.83% and 43.48% samples from condensed matter physics and high energy physics tend to be misclassified as astrophysics, respectively, which is attributed to the imbalanced dataset. **Figure 3g** shows the breakdown of MAC operations in graph learning and compares the energy consumption of random resistor array-based hardware with that of a digital system. Similar to experiments on MUTAG molecular classification, the majority of MACs are contributed by graph embedding procedure, leading to an overall energy consumption ~3.72 mJ per forward propagation of the entire dataset, considerably lower than that of conventional implementation, i.e., ~346.20 mJ, achieving 93.2× improvement in inference energy efficiency (see Method and Supplementary Figure 7 for the impact of hyperparameters on system energy efficiency). In addition, the number of MAC operations (energy) for optimizing the ESGNN is ~1.16 MOPs (~23.49 μJ) compared to ~16.98 GOPs (~0.34 J) of one epoch SGD training of a graph neural network with the same amount of parameters thanks to the fixed and random weights of the ESGNN, effectively reducing the total training cost by ~99.46% (see Supplementary Note 1).

# Node classification using ESGNN

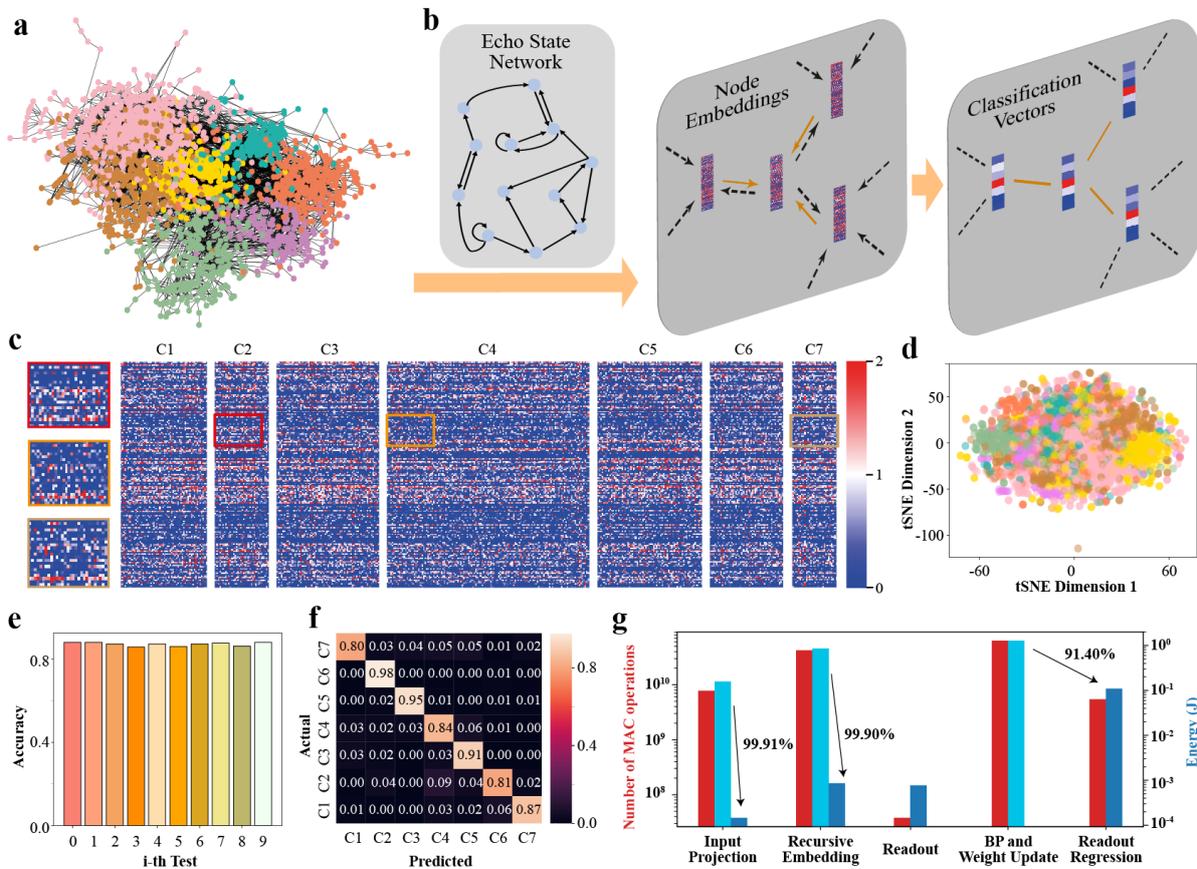

Figure 4 | **Node classification of a citation network. a**, Illustration of the large-scale citation network CORA. Each node in the graph is a scholarly article while an edge indicates a citation between two papers. There are a total of 7 categories of the articles, indicated by node colours, according to their disciplines. **b**, Node classification scheme. The input graph is first embedded using the ESGNN according to the protocol shown in **Figure 1f** and Methods, followed by a graph convolution layer serving as the readout to produce a classification vector for each node. **c**, Illustration of simulated node embeddings. Coloured boxes on the left are the zoom-in of node embedding details. **d**, Node embedding mapped to a 2D spacing using t-distributed stochastic neighbour embedding (tSNE), showing clear clustering of nodes of the same categories. **e**, Accuracy of 10 random tests for node classification. The average accuracy is 87.12%, comparable to state-of-the-art algorithms. **f**, Normalized confusion matrices of the simulated classification results. **g**, Breakdown of estimated MAC operations (red bars) and associated energy consumption (light blue bars for the conventional digital system and dark blue bars for our system). In forward (backward) propagation, the conventional digital hardware and our random resistor array-based system consumes ~1.01 J (~1.28 J) and ~1.77 mJ (~109.87 mJ), respectively, revealing >500× improvement in the inference

energy efficiency (and ~95.12% reduction of the total training energy cost).

Despite the graph classification tasks, node classification tasks constitute another important category of graph learning. We simulate our ESGNN in solving a large-scale node classification problem with the CORA citation network dataset[67], which is schematically illustrated in **Figure 4a**. This graph contains 2708 nodes, each of them represents a scientific publication and belongs to one of the seven research disciplines labelled by the node colour. Each edge of the graph represents a citation relationship between two publications. The input node features are 1433-dimensional word vectors. The graph is then fed into the ESGNN for node embeddings, as shown in **Figure 4b**. Different from the graph classification tasks which employ a trainable fully connected readout layer, a single software graph convolution layer serves as the readout layer with trainable weights to classify nodes, improving accuracy at low hardware and time cost (see Methods). **Figure 4c** shows the node embeddings of the whole dataset according to **Figure 1f**, grouped by node classes, where some dimensions are highly discriminative across different classes. **Figure 4d** shows the distribution of node embedding in a 2D space using t-distributed stochastic neighbour embedding (tSNE) dimension reduction. Nodes from the same category are clearly clustered without any supervision (see Supplementary Figure 4 for visualizing node embedding distributions in 3D space). To evaluate the performance, we measure the accuracy of ESGNN with different randomly initialized weights (see Methods). The 10-time average test accuracy reaches 87.12% in **Figure 4e**, which is comparable to those of state-of-the-art algorithms such as GCN (86.64%)[3] and GAT (88.65%)[4] running on conventional digital systems (see Supplementary Figure 6 for the impact of hyperparameters on accuracy). **Figure 4f** shows the simulated confusion matrix that is dominated by diagonal elements, affirming the high classification accuracy. To benchmark the efficiency of our random resistor arrays in solving this node classification problem, we count the MAC operations of different steps, as shown by the red bars in **Figure 4g**. The total number of operations is 50.02 GOPs per forward pass of the entire dataset, where the majority of them comes from multiplications with the recurrent weight matrix $W_R$, *i.e.* 42.22 GOPs, while the second largest is from multiplications with the input weight matrix $W_I$, *i.e.* 7.76 GOPs. The corresponding energy consumptions for inference are ~1.10 J and ~1.17 mJ for the conventional digital system (light blue bars) and our system (dark blue bars), respectively, affirming the large boost of energy efficiency in node classification (see Supplementary Figure 7 for the hyperparameters impact on operations and energy). In addition, the number of MAC operations and corresponding energy consumption for training our ESGNN are ~5.43 GOPs and ~109.87 mJ per epoch, while that for a graph neural network of the same number of parameters is ~63.18 GOPs and ~1.28 J per epoch respectively, indicating a 95.12% reduction of the total training cost (see Supplementary Note 1).

## Discussion

In this paper, we demonstrate a hardware-software co-design scheme for graph learning. Hardware-wise, the stochasticity of resistive switching is leveraged to produce low-cost and scalable random resistor arrays which physically implement the weights of an ESGNN, featuring in-memory computing with large parallelism and high efficiency that overcomes the von Neumann bottleneck. Software-wise, ESGNN not only takes the advantage of physical random projections enabled by random resistor arrays in performing graph embedding, but also substantially reduces the training cost of traditional graph learning. The resultant system demonstrates the great potential as a brand-new edge learning platform for graphs.

**Methods**

**Fabrication of resistive memory chips.** The resistive memory chip consists of a 512×512 crossbar array. Each of the resistive memory cells is integrated on the 40 nm standard logic platform. The resistive memory cells, including bottom electrodes, top electrodes, and transition-metal oxide dielectric layer are built between Metal 4 and Metal 5 layers of the back-end-of-line process. The via of bottom electrodes, with a diameter of 60 nm, is patterned by photolithography and etching. The via is filled by TaN by physical vapor deposition followed by chemical mechanical polishing. A buffer layer of 10 nm TaN is deposited by physical vapor deposition on the bottom electrode via. Then 5 nm Ta is deposited and then oxidized in an oxygen ambident to form an 8 nm $TaO_x$ dielectric layer. The top electrodes comprise 3 nm Ta and 40 nm TiN, which are sequentially deposited by physical vapor deposition. After fabrication, the logic back-end-of-line metal is deposited using the standard logic process. The cells in the same columns share top electrode connections while those of in the same rows share bottom electrode connections. Finally, the chip has been post-annealed in vacuum at 400°C for 30 minutes.

**The hybrid analogue-digital computing platform.** As shown in Supplementary Figure 1, the platform consists of an 8-channel digital-to-analogue converter (DAC80508, TEXAS INSTRUMENTS, 16-bit resolution) and two 8-bit shift registers (SN74HC595, TEXAS INSTRUMENTS) to source 64-ways parallel analogue voltages with 8-independent voltage amplitudes in the range from 0 V to 5 V. To perform vector-matrix multiplication, a DC voltage is applied to bit lines of the resistive memory chip through a 4-channel analogue multiplexer (MUX) (CD4051B, TEXAS INSTRUMENTS). The results are represented by currents from source lines and converted to voltages by trans-impedance amplifiers (OPA4322-Q1, TEXAS INSTRUMENTS). The voltages are then read by an analogue-to-digital converter (ADS8324, TEXAS INSTRUMENTS, 14-bit resolution) which passes the readings to the Xilinx SoC.

**Multibit vector-matrix multiplication.** To perform vector-matrix multiplication, the analogue input vector is first digitized into an *m*-bit binary vector where each element is an *m*-bit binary number. ($m = 4$ in this case) The analogue multiplication is therefore approximated by *m* times multiplication with binary input vectors corresponding to different significance. In each multiplication, a row is biased to a small fixed voltage (*e.g.* 0.3 V) if it receives a bit "1" or grounded if it receives a bit "0". The output currents of all columns are acquired sequentially using the column MUX. The resultant currents are multiplied with the significance and accumulated in the digital domain. Note that a larger *m* leads to better precision but an increased cost of energy and time.

**Graph classification experiments.** As shown in **Figure 1c**, the crossbar array is partitioned logically into two conductance matrices $\boldsymbol{G_I} \in \mathbb{R}^{h \times (u+1)}$ and $\boldsymbol{G_R} \in \mathbb{R}^{h \times h}$ (*u* and *h* represent the dimension of input feature vector and the number of hidden neurons, $h = 50$ for both the MUTAG and COLLAB datasets) which are then mapped to the input weight matrix $\boldsymbol{W_I} = \alpha_I \boldsymbol{G_I} \in \mathbb{R}^{h \times (u+1)}$ and recursive matrix $\boldsymbol{W_R} = \alpha_R \boldsymbol{G_R} \in \mathbb{R}^{h \times h}$, respectively. Here, scaling factors $\alpha_I$ and $\alpha_R$ are hyperparameters, which are set to 0.0016/μS and 0.006/μS, respectively in the graph classification experiments to ensure the echo state properties of the network (spectral radius less than unity).

Given a graph $\mathcal{G} = (\mathcal{V}, \mathcal{E})$ with *n* nodes $\boldsymbol{x_i} \in \mathcal{V}$ and edges $(\boldsymbol{x_i}, \boldsymbol{x_j}) \in \mathcal{E}$, we first compute the input projection of each node feature vector. For *j*-th node, its node input feature vector is $\boldsymbol{x_j} \in \mathbb{R}^{u+1}$ (with a unit bias) and the input projection is $\boldsymbol{u_j} = \boldsymbol{W_I}\boldsymbol{x_j} \in \mathbb{R}^h$, where $\boldsymbol{x_j}$ is quantized to a 4-bit binary vector mapped to voltages applied to the random resistor array. For the MUTAG dataset, the node feature vectors are the concatenations of one-hot vectors and the bias ($\boldsymbol{x_j} \in \mathbb{R}^{7+1}$), denoting their atom types. While for the COLLAB dataset, the node feature vectors are constant, that is the concatenation of a unit scalar and the bias ($\boldsymbol{x_j} \in \mathbb{R}^{1+1}$). The node internal state vector, or its embedding, is then iteratively updated. The internal state vector of $j$-th node at time $t+1$, denoted by $s_j^{(t+1)} \in \mathbb{R}^h$, is computed by aggregating its state $s_j^{(t)}$, input projection $u_j$, and the states of all its neighbors after random projections by the recursive matrix $\sum_{k \in N(j)} W_R s_k^{(t)}$ (here $N(j)$ denotes the set of neighbouring nodes of node $j$) according to Eqn. (1).

$$s_j^{(t+1)} = a s_j^{(t)} + (1-a)\sigma\left[u_j + \sum_{k \in N(j)} W_R s_k^{(t)}\right], \qquad (1)$$

where $\sigma$ is the activation function (*tanh*, in this work), $a$ is the leaky factor (0.2, in this work) and $s_k^{(t)}$ is quantized to a 4-bit binary vector mapped to voltages applied to the random resistor array. All other arithmetic operations are performed in the digital domain.

The graph embedding is computed by sum pooling of all node embeddings of a given graph to extract a single feature vector as the representation of the graph, or mathematically $g = \sum_j s_j^{(T)} \in \mathbb{R}^h$, where $T$ is the total number of iterations. Unlike classical echo state networks, the node internal state in ESGNN iterates finite times (see Supplementary Figure 6), as a trade-off between accuracy, energy cost, and over-smoothing.

The readout layer is a fully connected layer implemented in the digital domain. For the MUTAG (COLLAB) dataset consisting of two (three) categories, the readout layer maps graph embedding vectors $g$ onto class vectors $o \in \mathbb{R}^2$ ($o \in \mathbb{R}^3$) using 102 (153) floating-point weights with bias. It shall be noted the 2-category classification can also be performed by mapping $g$ onto class scalars $o \in \mathbb{R}$ to further reduce the number of weights of the readout layer. During training, we first evaluate graph embeddings of the entire training set. The embeddings and the labels are then concatenated for evaluating the weights of the fully connected readout layer using linear regression.

All hyperparameters (*e.g.* weight scaling factors $\alpha_I$ and $\alpha_R$, iteration time $T$, leaky rate $a$) are optimized by grid searching the hyperparameter space to maximize the hardware performance in the 10-fold cross-validation tests.

**Node classification simulation.** For CORA simulation, we use PyTorch 1.9.0 as the deep learning framework and Torch-geometric 1.7.2 as the graph deep learning tool. The CORA dataset visualized in **Figure 4a** uses the force-directed Kamada-Kawai algorithm, where the data are grouped by classes. The coordinates of nodes have been slightly refined for better visualization. The node embedding follows the same protocol with that of graph classification tasks using 1000 neurons. The readout layer is a single graph convolutional layer. During the training, the readout layer is optimized using stochastic gradient descent by minimizing a cross-entropy loss function. The readout layer has been trained for 200 epochs with a learning rate 0.01, a weight decay factor 0.005, a momentum 0.9, and a dropout rate 0.2. The performance of the model is assessed by training the readout layer upon different randomly initialized weights (10 sets of weights were used here).

**Acknowledgements**

This research is supported by the National Key R&D Program of China (Grant No. 2018YFA0701500), the National Natural Science Foundation of China (Grant Nos. 61874138, 61888102, 61821091), the Strategic Priority Research Program of the Chinese Academy of Sciences (Grant No. XDB44000000), Hong Kong Research Grant Council - Early Career Scheme (Grant No. 27206321), National Natural Science Foundation of China - Excellent Young Scientists Fund (Hong Kong and Macau) (Grant No. 62122004). This research is also partially supported by ACCESS – AI Chip Center for Emerging Smart Systems, sponsored by Innovation and Technology Fund (ITF), Hong Kong SAR.